# Metaclasses and Reflection in Smalltalk

A. N. Clark, Department of Computing, University of Bradford
Bradford, West Yorkshire, BD7 1DP, UK
e-mail: A.N.Clark@comp.brad.ac.uk, tel.: (0274) 385133
September 18, 1997

## 1  Abstract

Many Object-Oriented Programming Languages provide reflective features which may be used to control the interpretive mechanism of the language. Often, these features are defined with respect to a *golden braid* consisting of objects, classes and metaclasses. This paper describes the Smalltalk golden braid and generalize it for multiple inheritance. Multiple inheritance leads to choices between many different inheritance strategies. The reflective features of Smalltalk cannot affect the basic mechanisms of inheritance and so an arbitrary choice must be made for multiple inheritance. A language is described in which the reflective features of Smalltalk are extended so as to allow programmer defined inheritance strategies.

## 2  Introduction

The evaluation of a programming language expression $e$ in a given context $c$ may be described by the evaluation of a program $p$ which takes a representation of $e$ and $c$ as input. $e$ is termed an object-level construct whilst $p$ and the representations of $e$ and $c$ are termed meta-level constructs. For illustration we use an operator $\mathcal{M}$ which maps object-level constructs to meta-level constructs. If the languages which are used for both the object- and meta-levels are the same and causally connected, then the language is *reflective* [27].

Object-Oriented Programming Languages (OOPLs) have interpretive mechanisms which are based upon classes, object creation, message passing and inheritance. Classes typically define the local state and operations for objects which are their instances. When a message is passed to an object, the operation with the message name is invoked with respect to the local state. A class inherits from another class by including all the inherited storage and operation definitions along with its own. The meta-level of an OOPL describes how to perform inheritance, message passing, instance creation *etc.* If the OOPL is reflective then these mechanisms are described in terms of messages which are sent to objects at the meta-level. Consider the objects at some base level $B$, the objects and messages which describe how to perform creation of objects at $B$, message passing at $B$ *etc.* are defined at level $\mathcal{M}(B)$; these objects are called classes. A class at level $\mathcal{M}(B)$ is characterized by controlling the creation and subsequent behaviour of a collection of objects at level $B$. The objects (which are classes) at level $\mathcal{M}(B)$ are created and controlled by objects at level $\mathcal{M}^2(B)$; these objects are called metaclasses. A metaclass at level $\mathcal{M}^2(B)$ is characterized by controlling the creation and subsequent behaviour of classes at level $\mathcal{M}(B)$.

[4] coined the term *golden braid* to describe the relationship between objects, classes and metaclasses. Of course since classes are objects then we can view the golden braid starting at

$\mathcal{M}(B)$ and ending at $\mathcal{M}^3(B)$ which reinterprets the metaclasses at level $\mathcal{M}^2(B)$ as classes.

A feature of an OOPL is builtin at all levels when its behaviour is the same for $\mathcal{M}^n, n \geq 0$. Such features cannot be extended and may be said to be *intransigent*. An OOPL can be reflective whilst still having intransigent features; an OOPL which has no intransigent features may be said to be *fully* reflective.

Reflective OOPLs include Smalltalk [19], Loops [4], CLOS [3] [24], KRS [34], ObjVLisp [13] [5] [6]. These languages differ in terms of the ways in which the golden braid is implemented and the extent to which it may be used to affect the basic interpretive mechanisms of the respective languages.

This paper describes the reflective power of Smalltalk, identifies an intransigent language feature (*send*) and proposes a language extension which increases the reflective power by including the feature into the meta-level. §3 describes a simple functional language which will be used to implement three reflective object-oriented systems which differ with respect to how expressive their reflective features are. §4 describes a system called Abstract Smalltalk (AS) which is an implementation of the relevant features of the language Smalltalk. AS exhibits single inheritance which is generalised to multiple inheritance in the system AS with Multiple Inheritance (ASMI) described in §5. Both AS and ASMI have intransigent language features including object representation and message passing which means that it is difficult to have multiple types of inheritance mechanism co-existsing in the same system. This is particularly a drawback when the system exhibits multiple inheritance because there are many orthogonal multiple inheritance schemes. ASMI with Reflective Send (ASMIRS), described in §6, makes both object representation and message passing a reflective feature of the system. This is shown to support different types of multiple inheritance strategy within the same system.

## 3 Functional Representation

The systems AS, ASMI and ASMIRS, are constructed using a simple functional language which, following [26], has been enriched with data values, operators and evaluation rules which are characteristic of object-oriented programming languages. The language has a call-by-value semantics [32] and enriches the evaluation rules for the $\lambda$-calculus with pattern matching, currying, first class environments and updateable locations. The syntax of the language is divided into two: the kernel syntax which is given a semantics using a state transition system based on the SECD machine [25], and the sugared syntax which is given a semantics by translating into the kernel. For more information about functional languages see [16] and [2]. The sugared syntax is given below:

$T ::= \textbf{let } D$
$D ::= I = E \mid F$
$F ::= I\ P^+ = E \mid \textbf{meth } I\ P^+ = E \mid F|F$
$P ::= I \mid \_ \mid (P,\ldots,P) \mid KP \mid N \mid S$
$E ::= I \mid N \mid S \mid \lambda P^+.E \mid EE \mid EOE \mid \textbf{if } E \textbf{ then } E \textbf{ else } E \mid (E,\ldots,E) \mid (E) \mid E;E$
$\qquad [E,\ldots,E] \mid E \textbf{ where } D^+ \mid \textbf{let } D^+ \textbf{ in } E \mid \textbf{case } E \textbf{ of } A \textbf{ end} \mid \textbf{open } E \textbf{ in } E$
$A ::= P \Rightarrow E \mid AA$

A program is a sequence of top level recursive definitions $t \in T$ and expressions $e \in E$. $D$ is the syntax of declarations which may be a simple value such as $i = 10$ or a functional declaration $f \in F$ such as $add(x,y) = x+y$. Functional declarations may be overloaded and may include methods, both of which are described in appendix A. Patterns $p \in P$ are used in binding positions to limit the domain of a function and to decompose the value which is supplied as an argument to the function by extracting sub-components and binding them to identifiers. A pattern may be an identifier $i$ in which case the supplied value is bound to $i$; a wildcard _ in which case the supplied value is ignored; a tuple $(p_1, \ldots, p_n)$ in which case the supplied value must be a tuple of the same length and the corresponding sub-components must match against the sub-patterns; a constructor $k \in K$ applied to a pattern $p$ in which case the supplied value must be constructed using $k$, e.g. $k(v)$, and $v$ must match $p$; or a constant number $n \in N$ or string $s \in S$ in which case the supplied value must be the constant.

An expression $e$ is an identifier, number or string; a function $\lambda p_1 p_2 \ldots p_n.e$ which may be curried and have patterns in the binding positions; a prefix application $e_1 e_2$; an infix application $e_1 \oplus e_2$ where $O$ defines a collection of infix operators; a conditional expression; a tuple; a parenthesized expression; a sequenced expression $e_1; e_2$ which is used to control side effects; a list expression $[e_1, \ldots, e_n]$ where $[e]$ constructs singleton lists and $[]$ is the empty list; a where or let expression both of which may have a sequence of declarations which are established in parallel; a case expression **case** $e$ **in** $p_1 \Rightarrow e_1 \ldots p_n \Rightarrow e_n$ **end** where $e$ is evaluated and tested against the patterns in turn, the first pattern which matches will deconstruct the value of $e$, possibly bind some identifiers, and evaluate the corresponding expression; an open expression **open** $e_1$ **in** $e_2$ where $e_1$ produces an environment binding identifiers to values which is added to the current environment for the scope of the evaluation of expression $e_2$.

Environments are collections of associations (bindings) between keys and values. The operational semantics of programming languages often uses environments which bind names to values in order to describe the identifiers which may be legally referenced at any point in the program execution and their respective values. Environments are a convenient representation for objects and the functional language provides builtin operators for constructing and manipulating environments as data values. The empty environment which binds no keys to values is the value $\{\}$, a singleton environment binding the key $k$ to the value $v$ is constructed using an infix operator $k \mapsto v$, a pair of environments are concatenated using an infix operator $e_1 \oplus e_2$. The value associated with a key in an environment is "looked up" using an infix operator $e \bullet k$, where the environment $e$ binds $k$ more than once, the rightmost value is returned. When $e$ does not bind $k$, the distinguished value $\epsilon$ is returned.

Functions are associated with environments which define the values for the identifiers which are freely referenced in the function bodies. The environment which is associated with a function (or *closure*) is returned using the *reification* operator $R$. The environment associated with a function is updated (producing a new function) using the *installation* operator $I$, which is applied to a pair $(e, f)$. Using environment primitives $R$ and $I$, the infix operator $\_ \hookrightarrow \_$ is defined which extends the environment of a function on the right:

$$\textbf{let } e \hookrightarrow f = I((R(f)) \oplus e, f)$$

The systems AS, ASMI and ASMIRS, implement objects using side effects. The builtin operator $\_ := \_$ evaluates its left hand operand to produce an updateable location (*i.e.* the address of the value in the state machine) which is shared with many other values in the program state. The contents of the location are updated to by the value of the right hand operand. All values in environments are updateable locations.

Cyclic data values are constructed using the paradoxical operator $\mathbf{Y}$ which finds the fixed

point of a function:
$$\mathbf{Y}(f) = f(\mathbf{Y}(f))$$

Lists are built using the constructors $[]$, $\_ :: \_$ and $\_ +\!\!+ \_$, list homomorphisms are constructed using the operators $\backslash_l$ and $\backslash_r$ which are defined as follows:

$$\backslash_r(\otimes)(f)(v)(v_1 :: (v_2 :: (\ldots :: (v_n :: [])))) = f(v_1) \otimes (f(v_2) \otimes \ldots \otimes (f(v_n) \otimes v))$$

$$\backslash_l(\otimes)(f)(v)(v_1 :: (v_2 :: (\ldots :: (v_n :: [])))) = (((v \otimes f(v_1)) \otimes f(v_2)) \otimes \ldots) \otimes f(v_n)$$

Environments are built using the constructors $\{\}$, $\_ \mapsto \_$ and $\_ \oplus \_$, environment homomorphisms are constructed using using the operator $/$, for example:

$$/(\otimes)(\odot)(v)((k_1 \mapsto v_1) \oplus \{\} \oplus (k_2 \mapsto v_2)) = (k_1 \odot v_1) \otimes v \otimes (k_2 \odot v_2)$$

Sets are used to implement objects in both ASMI and ASMIRS. The empty set is $\phi$, $\{\_\}$ constructs singleton sets, $\_\cup\_$ is the set union operator and $\_-\_$ is the set difference operator. The functional language supports set comprehensions, it is beyond the scope of this paper to describe how such comprehensions are implemented in functional languages – see [35] for more details.

The operator *splitlistl* will map a pair $(v, l)$ to a pair $(l_1 +\!\!+ [v], l_2)$ such that $l = l_1 +\!\!+ [v] +\!\!+ l_2$ and $v$ occurs only once in $l$. The operator *splitlistr* is similiar except the result is $(l_1, [v] +\!\!+ l_2)$. The operator *find* is applied to a predicate $p$, a value $v$ and a list $l$ and will return the first value in $l$ which satisfies the predicate $p$ or $v$ otherwise.

The semantics of the functional language is described by a translation to the kernel language in appendix A.

## 4 Abstract Smalltalk

Smalltalk implements a golden braid with both classes and metaclasses treated as proper objects. The metaclasses are used in a slightly restricted way such that a metaclass has only one instance. Smalltalk is not fully reflective since the representation of objects and the semantics of message passing are intransigent language features. This section describes the Smalltalk golden braid in terms of a system called Abstract Smalltalk (AS). §4.1 describes the basic classes necessary to support the Smalltalk interpretive mechanism, §4.2 describes how AS is represented using the language of §3, §4.3 describes the main AS operations of object creation and message passing and §4.4 gives a meta-circular definition of AS.

### 4.1 Basic Classes

An AS *object* is a data value which contains two environments: an *instance variable* environment and a *method* environment. The method environment binds names to functions each of which can be invoked by "sending the object a message" containing the name (referred to as the message *selector*) and the actual parameter. The instance variable environment provides bindings for identifiers which may be referenced in the body of the methods, but otherwise cannot be accessed.

Each object is created by instantiating a *class* which defines the method environment and the names which will be bound in the instance variable environment. The instantiation process supplies values for each of the instance variable names. A class is created as a *subclass* of another class which is referred to as its *superclass*, the new class will contain all of the instance variable names and method definitions from its superclass – this is termed *inheritance*.

In AS, all classes are also objects. A class is distinuished from an ordinary object because its method environment contains a function which can create new objects and its instance variable environment contains the methods and instance variable names which all instances of the class will contain.

Classes are instances of classes which are termed *metaclasses*. A metaclass is distinguished from an ordinary class because its instance variable environment contains definitions of the instance variables and methods which are necessary to be a class. As is shown in §4.4, it is not necessary to define the term *metametaclass* because metaclasses can be conveniently defined *in terms of themselves*.

AS is defined as a collection of basic classes each of which has an instance variable environment and a method environment:

$$ienv = (\text{``ivars''} \mapsto [\text{``}x\text{''}, \text{``}y\text{''}]) \oplus (\text{``menv''} \mapsto (\text{``init''} \mapsto pointinit))$$
$$menv = (\text{``new''} \mapsto cdnew) \oplus (\text{``init''} \mapsto cdinit)$$

which is the instance variable and method environments for a class which creates two dimensional points. Each instance of this class will have two instance variables, namely $x$ and $y$, which are initialised using the method named *init*, implemented as the function *pointinit*. An instance is created by sending the class a *new* message which is implemented by the function *cdnew*. The class itself was initialised using the method named *init* which is implemented using the function *cdinit*. An instance of this class is the following:

$$ienv = (\text{``}x\text{''} \mapsto 10) \oplus (\text{``}y\text{''} \mapsto 100)$$
$$menv = (\text{``init''} \mapsto pointinit)$$

The basic classes which are necessary to support the AS system are shown in figure 1 they are: *object*, which all other classes inherit from, described in §4.1.1; *cd*, which is a metaclass describing the minimum amount of information necessary to be a class, described in §4.1.2; *mc*, which is a metaclass describing the minimum amound of information necessary to be a metaclass, described in §4.1.3; finally, *class* which is a metaclass describing the minimum amount of information necessary to be a non-metaclass class, described in §4.1.4.

Figure 1 can be interpreted by "chasing links". In order to find out what instance variables and methods an object has, first follow the instance link to the class which was used to create the object. This class will define a collection of instance variable names which are bound to values in the *ienv* component of the object and a collection of methods which form part of the *menv* component of the object. Next, follow the superclass link to a class which defines further variable names and methods which are found in the *ienv* and *menv* components of the object respectively. Continue this process until *object* is reached. All superclass links eventually lead to *object* which is referred to as the root of the AS object inheritance tree. All classes form the root of an inheritance tree which identifies a collection of objects which contain the variable names and methods defined by the class.

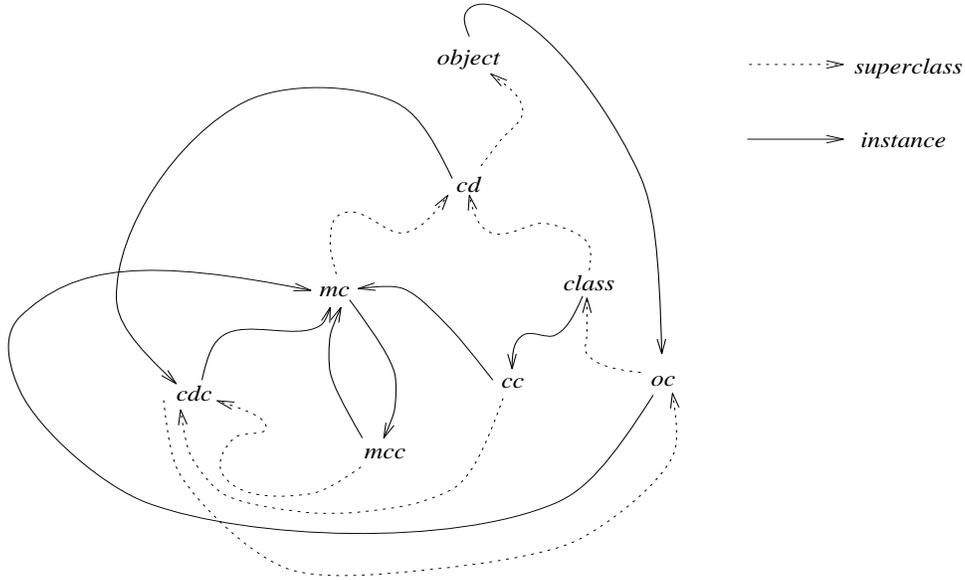

Figure 1: Initial AS class configuration

### 4.1.1 Object

$$
\begin{aligned}
ienv = &(\text{``super''} \mapsto nullclass) \oplus \\
&(\text{``ivars''} \mapsto [\text{``class''}]) \oplus \\
&(\text{``menv''} \mapsto (\text{``init''} \mapsto objinit) \oplus \\
&\qquad\qquad (\text{``dnu''} \mapsto objdnu)) \oplus \\
&(\text{``class''} \mapsto oc)
\end{aligned}
\qquad
\begin{aligned}
menv = &(\text{``new''} \mapsto cdnew) \oplus \\
&(\text{``init''} \mapsto cdinit) \oplus \\
&(\text{``subclass''} \mapsto classsub) \oplus \\
&(\text{``dnu''} \mapsto objdnu)
\end{aligned}
$$

The class *object* is the root of the AS inheritance tree and as such defines the minimum instance variable names and methods in order to be an object. The superclass of *object* is the pseudo class *nullclass* which is necessary to allow all classes to uniformly inherit from somewhere. *object* defines a single instance variable named *class* which is inherited by all AS classes and therefore will occur in the *ienv* component of all AS objects. The value of this variable in each object will be the class which was used to create the object. *object* defines two methods which will be inherited by all AS classes and therefore will occur in the *menv* component of all AS objects (unless shadowed by a subclass definition). The method named *init* is implemented using the function *objinit* and is used to initialise the object after it is created. The single purpose of *obnjinit* is to set the *class* variable. The method named *dnu* is implemented using the function *objdnu* and is invoked when a message is sent to an object for which there is no *menv* component.

The *menv* component of *object* contains the methods which define the behaviour of *object*. This is the standard class behaviour: *new* is used to create an instance of the class, *init* is used to initialise the class after it is created, *subclass* is used to create a class which is a subclass of *object* and *dnu* is invoked if *object* is ever sent a message whose name is not bound in *menv*.

### 4.1.2 Class Description

$$ienv = (\text{``super''} \mapsto object) \oplus$$
$$(\text{``ivars''} \mapsto [\text{``super''}, \text{``ivars''}, \text{``menv''}]) \oplus$$
$$(\text{``menv''} \mapsto (\text{``new''} \mapsto cdnew) \oplus$$
$$(\text{``init''} \mapsto cdinit)) \oplus$$
$$(\text{``class''} \mapsto cdc)$$

$$menv = (\text{``new''} \mapsto cdnew) \oplus$$
$$(\text{``init''} \mapsto cdinit) \oplus$$
$$(\text{``subclass''} \mapsto classsub) \oplus$$
$$(\text{``dnu''} \mapsto objdnu)$$

The class $cd$ is the root of the AS class inheritance tree and as such defines the minimum instance variable names and methods in order to be a class. The superclass of $cd$ is $object$, so $cd$ inherits the instance variable names (i.e. $class$) and the methods (i.e. $init$ and $dnu$) from $object$. The minimum instance variables which are necessary to be a class are: $super$ whose value must be a class whose instance variable names and methods are inherited; $ivars$ whose value must be a list of instance variable names; and $menv$ whose value must be an environment binding method names to functions. Notice that all the classes described in this section have $ienv$ components which bind variables $super$, $ivars$, $menv$ and $class$ which originate from $cd$ ($class$ being inherited from $object$).

### 4.1.3 Metaclass

$$ienv = (\text{``super''} \mapsto cd) \oplus$$
$$(\text{``ivars''} \mapsto []) \oplus$$
$$(\text{``menv''} \mapsto (\text{``subclass''} \mapsto metasub)) \oplus$$
$$(\text{``class''} \mapsto mcc)$$

$$menv = (\text{``new''} \mapsto cdnew) \oplus$$
$$(\text{``init''} \mapsto cdinit) \oplus$$
$$(\text{``subclass''} \mapsto classsub) \oplus$$
$$(\text{``dnu''} \mapsto objdnu)$$

The class $mc$ is the root of the AS metaclass inheritance tree and as such it defines the minimum instance variable names and methods in order to be a metaclass. Each time a class is constructed, an instance of $mc$ is also constructed as its unique metaclass. The metaclass will define the methods for its sole instance which are extensions to the standard class behaviour as defined by $cd$. Furthermore, when a class is sent a message telling it to create a subclass of itself, its metaclass is also sent a subclass message, so the metaclass inheritance tree corresponds exactly to the class inheritance tree. This is seen in figure 1 where the classes $object$, $mc$, $cd$ and $class$ are instances of the metaclasses $oc$, $mcc$, $cdc$ and $cc$ where the inheritance between the metaclasses follows exactly that between the corresponding classes.

The class $mc$ defines a method $subclass$ which is used to create a subclass of a given metaclass. The method is implemented using the function $metasub$ which will create an instance of $mc$ and initialise this to have the receiver of the message as its superclass.

### 4.1.4 Class

$$ienv = (\text{``super''} \mapsto cd) \oplus$$
$$(\text{``ivars''} \mapsto []) \oplus$$
$$(\text{``menv''} \mapsto (\text{``subclass''} \mapsto classsub)) \oplus$$
$$(\text{``class''} \mapsto cc)$$

$$menv = (\text{``new''} \mapsto cdnew) \oplus$$
$$(\text{``init''} \mapsto cdinit) \oplus$$
$$(\text{``subclass''} \mapsto classsub) \oplus$$
$$(\text{``dnu''} \mapsto objdnu)$$

The class $class$ is the root of the AS class (non-metaclass) inheritance tree and as such defined the minimum instance variable names and methods in order to be a (non-metaclass) class. $class$ defines a method $subclass$ which is used to create a subclass of the receiver. This method deals with constructing a new metaclass by sending a $subclass$ message to the class of the receiver and then creating an instance of this metaclass. *subclass* is implemented by the function *classsub*.

A *subclass* message is sent to a receiver $c_1$ as follows:

$$send(c_1, \text{``subclass''}, (e_1, l, e_2))$$

where $e_1$ is an environment of methods which are termed "class methods" and which will form part of the *menv* component of the result; $l$ and $e_2$ are a list of names and a method environment which are the instance variables and methods defined by the new subclass. Assuming that $c_1$ is a non-metaclass, then the *send* expression is equivalent to the following:

$$send(send(c_2, \text{``subclass''}, e_1), \text{``new''}, [c_1, l, e_2])$$

where $c_2$ is the class of $c_1$. Since $c_2$ is a metaclass it will handle the *subclass* message differently from $c_1$. Upon receiving a *subclass* message, $c_2$ will create an instance of *mc*, supplying the value of the superclass, instance variable names and method environment as $c_2$, [] and $e_1$ respectively:

$$send(send(mc, \text{``new''}, [c_2, [], e_1]), \text{``new''}, [c_1, l, e_2])$$

The result of this expression is a class:

$$\begin{aligned}
ienv = &(\text{``super''} \mapsto c_1) \oplus & menv = &e_1 \oplus \\
&(\text{``ivars''} \mapsto l) \oplus & &(\text{``new''} \mapsto cdnew) \oplus \\
&(\text{``menv''} \mapsto e_2) \oplus & &(\text{``init''} \mapsto cdinit) \oplus \\
&(\text{``class''} \mapsto c_3) & &(\text{``subclass''} \mapsto classsub) \oplus \\
& & &(\text{``dnu''} \mapsto objdnu)
\end{aligned}$$

where the class $c_3$ is:

$$\begin{aligned}
ienv = &(\text{``super''} \mapsto c_2) \oplus & menv = &(\text{``new''} \mapsto cdnew) \oplus \\
&(\text{``ivars''} \mapsto []) \oplus & &(\text{``init''} \mapsto cdinit) \oplus \\
&(\text{``menv''} \mapsto e_1) \oplus & &(\text{``subclass''} \mapsto metasub) \oplus \\
&(\text{``class''} \mapsto mc) & &(\text{``dnu''} \mapsto objdnu)
\end{aligned}$$

## 4.2 Representation

An object is either null, which corresponds to an instance of the superclass of *object*, or a basic object which corresponds to an instance of anything else. Classes are layered like onions where the outermost layer corresponds to the most recent subclass, the next inner layer to its superclass, the next inner layer to its supersuperclass *etc*. The heart of the onion is the value *nullclass* which is the pseudo superclass of *object*. An object mirrors the onion-like structure of its class with each layer containing the instance variables and methods declared by the corresponding layer of the class. Each layer of an object has an additional value which is the entire object and is referred to as *self*. The heart of an object-onion $o$ is the null object, $nullobj(o)$, which is a pseudo instance of *nullclass*.

An object is represented as $obj(e_1, e_2, o_1, o_2)$ where $e_1$ is an environment associating instance variable names with locations containing their values (*ienv*), $e_2$ is an environment associating method names with methods (*menv*), $o_1$ is an object which is the next innermost layer of the object-onion referred to as *super* and $o_2$ is the whole object-onion referred to as *self*.

## 4.3 Basic Operations

The AS system is defined as a collection of basic classes whose methods are implemented using functions *objdnu*, *cdnew*, *etc.*, and a message passing operator *send*. The functions which implement the class methods are defined in terms of a single operator whose job is to construct objects in a particular format. This operator, *mkobj*, is defined in §4.3.1. The operator *send* implements the message delivery service which searches an object for a named method and then invokes it. *send* is defined in §4.3.2.

Both object creation and message passing are defined in terms of the concatenation of the instance variable environments which are contained in an object. The operator *getallenv*, defined below, is used to construct this environment:

$$\textbf{let } getallenv(nullobj(\_)) = \{\} \mid$$
$$getallenv(obj(e,\_,o,\_)) = e \oplus (getallenv(o))$$

### 4.3.1 Object Creation

The operator *mkobj* defined below:

$$\textbf{let } mkobj(nullclass)(o) = nullobj(o) \mid$$
$$mkobj(c)(o) = \textbf{open } getallenv(c) \textbf{ in } obj(\backslash_r(\oplus)(\mapsto \gamma)(\{\})(ivars), menv, mkobj(super)(o), o)$$

This operator is the primitive AS object creation operator. The argument $c$ is a class which is to be instantiated and $o$ is an object which will be the *self* component of the resulting instance. A class will bind the variables *super*, *ivars* and *menv* in its *ienv* component. A new instance is created layer by layer. Each layer has the format $obj(e_1, e_2, o', o)$ where $e_1$ binds the instance variable names from the corresponding layer of the class $c$ to the null value $\gamma$, $e_2$ is the method environment from the corresponding layer of $c$ and $o'$ is the result of instantiating the superclass of $c$.

### 4.3.2 Message Passing

The operator *send* defined below:

$$\textbf{let } send(nullobj(o), n, v) = send(o, \text{``dnu''}, (n, v)) \mid$$
$$send(obj(e_1, e_2, o_1, o_2), n, v) =$$
$$\quad \textbf{if } n \in dom(e_2)$$
$$\quad \textbf{then } (e_2 \bullet n)(o_2, o_1, e_1 \oplus (getallenv(o_1)))(v)$$
$$\quad \textbf{else } send(o_1, n, v)$$

performs AS message passing. The operator has three parameters which correspond to the receiver of the message, the selector of the message and the actual parameter of the message. The operation of *send* can be described in terms of the object-onion model (described in §4.2). When a message is sent to an object, the methods defined by the outermost layer are searched for a method whose selector matches. If such a method is found then it is activated, by applying it to objects $o_2$ and $o_1$, an environment $e_1 \oplus (getallenv(o_1))$ and value $v$. $o_2$ is the whole object-onion no matter which particular layer contains the matching method; $o_1$ is the object-onion constructed by stripping away the layer which contains the matching method – if a message is sent to $o_1$ it will continue the search where *send* left off; $e_1 \oplus (getallenv(o_1))$ is the concatenation of all instance variable environments in all layers from that which contained the matching method to the heart of the object-onion; $v$ is the actual parameter of the message. If no method matches then the process continues with the next innermost onion-layer. If the process encounters the heart of the onion then no method was defined for the selector and "$dnu$" is sent to $o_2$.

## 4.4 Definition

AS is defined as an initial collection of classes and a delivery service, $send$. The classes define methods which are implemented using the functions which are given below:

**let meth** $objinit(c) = class := c; self$
**let meth** $objdnu(m) = error(\text{"message"} \mathbin{+\!+} str(m) \mathbin{+\!+} \text{"not understood"})$
**let meth** $cdinit([c, l, e] \mathbin{+\!+} v) = super := c; ivars := l; menv := e; send(next, \text{"init"}, v)$
**let meth** $cdnew(v) = send(\mathbf{Y}(mkobj(self)), \text{"init"}, v \mathbin{+\!+} [self])$
**let meth** $metasub(e) = send(mc, \text{"new"}, (self, [], e))$
**let meth** $classsub(e_1, l, e_2) = send(send(class, \text{"subclass"}, e_1), \text{"new"}, [self, l, e_2])$

Collections of methods and instance variable names are predefined for convenience[1]:

**let** $om = (\text{"init"} \mapsto objinit) \oplus (\text{"dnu"} \mapsto objdnu)$
**let** $cdm = (\text{"new"} \mapsto basicnew) \oplus (\text{"init"} \mapsto classinit)$
**let** $cdv = [\text{"super"}, \text{"ivars"}, \text{"menv"}]$
**let** $mm = \text{"subclass"} \mapsto metasub$
**let** $cm = \text{"subclass"} \mapsto classsub$

The definitions of the AS classes are given below:

**let** $object = send(oc, \text{"new"}, [nullclass, [\text{"class"}], om])$
$oc = send(mc, \text{"new"}, [class, [], \{\}])$
$cd = send(object, \text{"subclass"}, (\{\}, cdv, cdm))$
$mc = send(cd, \text{"subclass"}, (\{\}, [], mm))$
$class = send(cd, \text{"subclass"}, (\{\}, [], cm))$

These definitions are meta-circular, *i.e.* the classes are assumed to exist for the process of their own definition. It is necessary to "bootstrap" AS by providing initial values for the classes which can be used to construct themselves. Using the description of chasing links given in §4.1, it is possible to determine the outcome of the meta-circular definitions. For example, it is possible to replace all messages with the selector "*subclass*" with a selector "*new*" by observing that all the receivers in question will inherit the method from *class* where it is implemented using *classsub*. The definitions have been partially expanded by chasing

---

[1]We should be careful about sharing and side effects here since the locations which are created in the predefined environments will occur more than once in the final system. A complete discussion of the issues of sharing is outside the scope of this paper and we define that no location is ever implicitly copied *i.e.* all occurrences of bindings will share and be affected by a single side effect.

links in the following definitions:

**let** $object = obj(\{\}, \{\}, o_1, object)$
  **where** $o_1 = obj(\{\}, \text{``subclass''} \mapsto classsub, o_2, object)$
    **where** $o_2 = obj(e, cdm, o_3, object)$
      **where** $e = (\text{``super''} \mapsto nullclass) \oplus (\text{``ivars''} \mapsto [\text{``class''}]) \oplus (\text{``menv''} \mapsto om)$
        $o_3 = obj(\text{``class''} \mapsto oc, om, nullobj(object), object)$
  $oc = obj(\{\}, mcm, o_1, oc)$
    **where** $o_1 = obj(e, cdm, o_2, oc)$
      **where** $e = (\text{``super''} \mapsto cdc) \oplus (\text{``ivars''} \mapsto cdv) \oplus (\text{``menv''} \mapsto cdm)$
        $o_2 = obj(\text{``class''} \mapsto mc, om, nullobj(oc), oc)$
  $cd = send(object, \text{``subclass''}, (\{\}, cdv, cdm))$
  $mc = send(cd, \text{``subclass''}, (\{\}, [], mcm))$
  $class = send(cd, \text{``subclass''}, (\{\}, [], cm))$

## 5  AS with Multiple Inheritance

The AS system described in §4 provides single inheritance which means that each class is allowed only one superclass whose state variable declarations and method definitions are inherited. This leads to a tree structure of classes with *object* as the root of the tree and new classes being added as leaves. This section will generalize AS to produce ASMI which is Abstract Smalltalk with Multiple Inheritance. Multiple inheritance allows each class to have more than one superclass from which it will inherit state variable declarations and method definitions. This leads to a graph structure of classes with the constraint that the graph must not contain cycles (*i.e.* a class cannot inherit from itself).

When classes are tree structured, as in AS, inheritance of variable declarations and methods is straightforward because there is no choice as to the order in which information from superclasses will be inherited. When classes are graph structured, as in ASMI, inheritance becomes more complex because there may be more than one route from one class to another. For example, since all classes inherit from *object*, any class which inherits from more than one superclass will construct at least two different inheritance paths from itself, though respective superclasses leading to *object*. Will the information from *object* be inherited twice? What happens when an object sends a message to *next*? Different programming languages offer different solutions to this problem, these include:

- Information is inherited as many times as it is reachable from the inheritor. This will lead to multiple copies of instance variable locations but has the advantage of being modular [33].

- Information is inherited only once. The graph is traversed from an inheritor, such that each node is encountered exactly once. This will cause no problems provided that the names of the methods are distinct. If they are not then a question arises with respect to which method will shadow the other and therefore which method will be executed when its selector is part of a message to *super*. One strategy is a left to right traversal of the inheritance graph, omitting a superclass if it has already been visited. Another strategy is a left to right traversal of the inheritance graph omitting a superclass if it will be visited later. Both strategies have the disadvantage that the shape of the inheritance graph is distorted with respect to inherited classes.

- Language facilities are provided so that all ambiguities which should arise are eliminated under programmer control [28].

It is not the purpose of this paper to discuss in depth the merits of various MI strategies (for more information see [7] [14] and [30]), for ASMI we will arbitrarily choose a left to right graph traversal which defers a superclass to its final occurence. ASMI and AS are compared in §5.1, the ASMI basic classes are described in §5.2, the representation of ASMI objects is defined in §5.3, the ASMI basic operations are defined in §5.4 and finally ASMI is defined in §5.5.

## 5.1 Comparison with AS

ASMI is a version of AS with multiple inheritance. The inheritance strategy which is adopted is that of a left to right graph traversal which defers shared superclasses to their final occurrence. Where each AS class has a single class as the value of its state variable "*super*", each ASMI class has a corresponding state variable called "*supers*" whose value is a list of classes. The empty list is used for the superclass of *object*, which delivers us from the irksome *nullclass*.

Each class forms the root of an acyclic directed graph where the nodes are classes and the edges represent inheritance links. In AS when a class is instantiated, the onion structure of the classes is translated to an identical onion structure for the instance in which all the instance variables are bound to initial values. In ASMI, when a class is instantiated, the graph structure of the classes is translated to an identical graph structure for the instance in which all the instance variables have been allocated new storage locations.

When an AS message is sent to an object, the search will progress through successive object layers until a method with the required selector is found. A subsequent message sent to *next* will continue the search until the heart of the onion structure is found. When an ASMI message is sent to an object, the search will perform a left to right depth first graph traversal until a method with the required selector is found. A subsequent message sent to *next* will continue the search until the final node in the graph is encountered.

In AS, a class is constructed by sending a *subclass* message to a class $c$ which will become the superclass of the new class. The class is created by sending a *subclass* message to the class of $c$, which creates a metaclass which is instantiated by sending it a *new* message. In ASMI, classes have multiple superclasses, so it is not possible to send a single *subclass* message without giving one of the superclasses an artificial status as the receiver of the message and all other superclasses lesser merit by supplying them as the argument. In ASMI, a class is created by sending a metaclass a message with the selector "*new*"; the superclasses of the new class are supplied as a list in the message argument. If a new metaclass is required, for example to define the class methods, then it must be created explicitly.

## 5.2 Basic Classes

Figure 2 shows the initial ASMI configuration of classes. Following ObjVLisp [13] it is not necessary for each class to have a single unique metaclass; the AS model whereby each class has a single unique metaclass (used to define the class methods) can be built from the general model upon which ASMI is based. In ASMI there are only two initial classes: *object* and *class*, all classes will have *object* as their ultimate superclass and all metaclasses will have *class* as their penultimate superclass. *object* describes the state variables and methods which are common to all classes and *class* describes the state variables and methods which are common to all metaclasses. *class* is an instance of itself and therefore represents the fixed point of the $\mathcal{M}$ operator discussed in §2. ASMI classes *object* and *class* are described in §5.2.1 and §5.2.2 respectively.

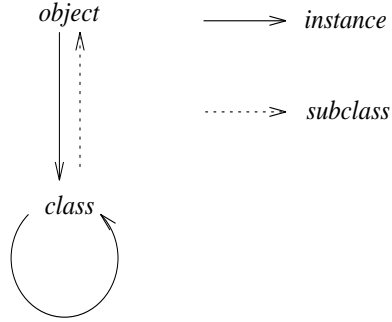

Figure 2: Initial ASMI class configuration

### 5.2.1 Object

$$
\begin{aligned}
ienv = &\ (\text{``supers''} \mapsto [\,]) \ \oplus \\
&\ (\text{``ivars''} \mapsto [\text{``class''}]) \ \oplus \\
&\ (\text{``menv''} \mapsto (\text{``init''} \mapsto objinit) \ \oplus \\
&\ \qquad\qquad (\text{``gc''} \mapsto objgc) \ \oplus \\
&\ \qquad\qquad (\text{``dnu''} \mapsto objdnu)) \ \oplus \\
&\ (\text{``class''} \mapsto class)
\end{aligned}
$$

$$
\begin{aligned}
menv = &\ (\text{``new''} \mapsto classnew) \ \oplus \\
&\ (\text{``init''} \mapsto classinit) \ \oplus \\
&\ (\text{``dnu''} \mapsto objdnu)
\end{aligned}
$$

*object* defines no superclasses, a single instance variable *class*, and three methods which initialise an object by setting the *class* instance variable, get the class of an object and handle the case when an object does not understand a message.

### 5.2.2 Class

$$
\begin{aligned}
ienv = &\ (\text{``supers''} \mapsto [object]) \ \oplus \\
&\ (\text{``ivars''} \mapsto [\text{``supers''}, \text{``ivars''}, \text{``menv''}]) \ \oplus \\
&\ (\text{``menv''} \mapsto (\text{``init''} \mapsto classinit) \ \oplus \\
&\ \qquad\qquad (\text{``new''} \mapsto classnew)) \\
&\ (\text{``class''} \mapsto class)
\end{aligned}
$$

$$
\begin{aligned}
menv = &\ (\text{``new''} \mapsto classnew) \ \oplus \\
&\ (\text{``init''} \mapsto classinit) \ \oplus \\
&\ (\text{``dnu''} \mapsto objdnu)
\end{aligned}
$$

*class* defines the instance variables *supers*, *ivars* and *menv*. These correspond to those of *cd* in AS except that for a given class, *supers* is a list of classes rather than a single class. *class* defines methods *new* and *init* which will instantiate and initialize the receiver repectively.

### 5.3 Representation

An ASMI object is a graph whose nodes contain instance variable and method environments and whose links represent inheritance. The representation of a graph is as follows:

$$g(e_1, S_1, e_2, e_3, S_2)$$

where $e_1$ is an environment mapping node addresses to nodes, $S_1$ is a set of edges, $e_2$ is an environment mapping edges to their source node address, $e_3$ is an environment mapping edges to their target node address and $S_2$ is a partial ordering on the nodes. A node is represented as

$$n(e_1, e_2)$$

where $e_1$ is an instance variable environment and $e_2$ is a method environment. The functions *getenv* and *getmenv* extract the instance variable and method environments from a node.

## 5.4 Basic Operations

Like AS, ASMI is defined in terms of a collection of basic classes, the functions which implement their methods and a message delivery service. In addition, objects are implemented as graphs and operations are provided which can be used to traverse graphs in various ways. §5.4.1 describes the operations for graph traversal, §5.4.2 describes basic object creation and §5.4.3 describes the ASMI message delivery service.

### 5.4.1 Graph Traversal

The basic operations which are used to traverse object graphs are defined as follows:

**let** $nullgraph = g(\{\}, \phi, \{\}, \{\}, \phi)$

**let** $gm(f)(g(e_1, S_1, e_2, e_3, S_2)) = g(/(\oplus)(\mapsto \circ (\mathbf{I} \times f))(\{\})(e_1), S_1, e_2, e_3, S_2)$

**let** $gmerge(g(e_1, S_1, e_2, e_3, S_2), g(e_4, S_3, e_5, e_6, S_4)) = g(e_1 \oplus e_4, S_1 \cup S_3, e_2 \oplus e_5, e_3 \oplus e_6, S_2 \cup S_4)$

**let** $root(g(e_1, S, e_2, e_3, \_)) = \{e_1 \bullet (e_2 \bullet e) \mid e \in S, \{e' \mid e' \bullet e_3 = e \bullet e_2\} = \phi\}$

**let** $targetnodes(n)(g(e_1, S, e_2, e_3, \_)) = \{e_1 \bullet (e_3 \bullet e) \mid e \in S, e_1 \bullet (e_2 \bullet e) = n\}$

**let** $walk(n)(g) = n :: (\backslash_r (+\!\!+)(walk)[](sort(targetnodes(n)(g))(ord(g))))$

**let** $traverse(g) = walk(n)(g)$ **where** $\{n\} = root(g)$

The operator *gm* is used to construct graph morphisms where $f$ is a function to be applied to all the nodes in the graph. The operator *gmerge* will merge two object graphs to produce a new object graph. $targetnodes(n)(g)$ produces the set of all nodes which are reachable from the node $n$ by traversing one edge in $g$. $root(g)$ is the set of all nodes which have no edges incident upon them (such sets will be singletons for properly formed objects). $traverse(g)$ is the sequence of nodes which are visited stating with the root of $g$ and visiting each node reachable from the root in an order which is consistent with both the ordering imposed by the edges of $g$ and the partial ordering component of $g$. $sort(S_1)(S_2)$ is the sequence of $S_1$ elements whose ordering is consistent with the partial ordering $S_2$.

The operator *traverse* will visit a node more than once if it is reachable using two or more paths from the root of a graph. ASMI will visit each node once when searching for a method in an object graph. The question arises as to which of the many paths to a node will be chosen. The two operators defined below, *onr* and *onl*, will order the nodes of an object graph so that nodes which can be encountered more than once are visited last and first respectively in a manner which is consistent with the ordering imposed by the graph edges and partial ordering. The operator *removeifmarked* will delete any nodes which have been marked in a special way and is explained in §5.4.3.

$$\begin{aligned} \textbf{let} \quad & l \otimes n = \textbf{if } n \in l \textbf{ then } l \textbf{ else } n :: l \mid \\ & n \otimes l = \textbf{if } n \in l \textbf{ then } l \textbf{ else } l +\!\!+ [n] \end{aligned}$$

$$\begin{aligned} \textbf{let} \quad & onr = removeifmarked \circ (\backslash_r (\otimes)(\mathbf{I})[]) \circ traverse \\ \textbf{let} \quad & onl = removeifmarked \circ (\backslash_l (\otimes)(\mathbf{I})[]) \circ traverse \end{aligned}$$

The instance variable environment of an object is extracted using the operator *getallenv*:

$$\textbf{let} \quad getallenv(o) = \backslash_r(\oplus)(getenv)(\{\})(onr(o))$$

### 5.4.2 Object Creation

A class may be viewed as two different graphs. The first is as an object graph since all classes are objects, where the nodes contain instance variable and method environments and the edges represent the inheritance of variables and methods. The second is as a class graph where the nodes contain instance variable names and method definitions relating to the instances of the class and edges represent inheritance of names and methods from the superclasses. Given a class graph and some initialisation values for the instance variables, instantiation is a simple graph morphism which retains the class graph structure and associates the values with the variable names in each node. The function *cg* defined below

$$\begin{aligned} \textbf{let} \quad & cg(g) = \\ & \textbf{let} \quad e = getallenv(g) \textbf{ in} \\ & \textbf{let} \quad n = n(e \bullet \text{``ivars''}, e \bullet \text{``menv''}) \textbf{ in} \\ & \textbf{let} \quad g = \backslash_r(gmerge)(cg)(nullgraph)(e \bullet \text{``supers''}) \textbf{ in} \\ & \textbf{in} \quad addnode(n, g) \end{aligned}$$

translates a class, represented as an object graph, to a class graph. $addnode(n, g)$ will add the node $n$ to the graph $g$ by allocating a new address for $n$ and will link $n$ to each node in $root(g)$ by allocating a new edge. The function *addnode* is not defined in this paper.

### 5.4.3 Message Passing

ASMI message passing is similar to AS message passing since the object is traversed to find a method with the given selector, but differs because the object representation is a graph and not a tree. ASMI methods have the same representation as AS methods, *i.e.* they have three hidden parameters for *self*, *next* and the instance variable environment. The value of the first parameter is the entire object graph. The value of the second parameter is the entire object graph, but the nodes which were traversed in order to find the method have been marked so that if a message is ever sent to *next* the marked nodes are ignored. Marking and unmarking of object graphs is performed by the following two functions:

$$\begin{aligned} \textbf{let} \quad & mark(g(n, e, s, t, o))(n') = g((n - \{n'\}) \cup \{mark(n')\}, e, s, t, o) \\ \textbf{let} \quad & unmark = gm(f) \textbf{ where } f(n(e_1, e_2) = n(e_1, e_2) \mid f(mark(n)) = n \end{aligned}$$

ASMI message delivery is performed by the function *send* which is defined below:

$$\begin{aligned}
\textbf{let}\ & send(o, n, v) = \\
& \textbf{let}\ x = \textit{find}(p)(\epsilon)(\textit{onr}(o))\ \textbf{where}\ p(\textit{node}) = n \in \textit{dom}(\textit{getmeths}(\textit{node})) \\
& \textbf{in if}\ x = \epsilon\ \textbf{then}\ \textit{send}(\textit{unmark}(o), \text{``}\textit{dnu}\text{''}, (n, v)) \\
& \quad \textbf{else let}\ o_1 = \backslash_l(\textit{mark})(\mathbf{I})(o)(\textit{1st}(\textit{splitlistl}(x, \textit{onr}(o)))) \\
& \qquad\qquad o_2 = \textit{unmark}(o) \\
& \qquad\qquad e = \backslash_r(\oplus)(\textit{getenv})(\{\})(\textit{2nd}(\textit{splitlistr}(x, \textit{onr}(o)))) \\
& \quad \textbf{in}\ ((\textit{getmeths}(x)) \bullet n)(o_2, o_1, e)(v)
\end{aligned}$$

The behaviour of *send* is as follows: *onr* is used to order the nodes in the object graph $o$ and $x$ is a node with a method environment containing the selector $n$. If no such $x$ is in $o$ then the message *dnu* is sent to $o$ after it is unmarked. Otherwise, $o_1$ is an object which will continue searching from node $x$ when it is sent a message. In order to continue from $x$, $o_1$ is produced by marking all of the nodes which have been traversed from the root node up to and including $x$. $o_2$ is an object which start searching from the original root node of $o$, so all marks are removed. $e$ is the environment constructed by concatenating all of the instance variable environments in the nodes which are traversed starting with $x$ and continuing on through the rest of $o$. Finally the method is applied to the hidden parameters $o_2$, $o_1$ and $e$ and the actual parameter, $v$, of the message.

## 5.5 Definition

ASMI is defined as an initial collection of classes and a delivery service *send*. The classes define methods which are implemented using the functions given below:

$$\begin{aligned}
& \textbf{let meth}\ objgc() = \textit{class} \\
& \textbf{let}\ \textit{instantiate}(c) = gm(\lambda(n(l,e)).\ n(\backslash_r(\oplus)(\mapsto \gamma)(\{\})(l),e))(cg(c)) \\
& \textbf{let meth}\ \textit{classnew}(v) = \textit{send}(\textit{instantiate}(\textit{self}), \text{``}\textit{init}\text{''}, v +\!\!+ [\textit{self}])
\end{aligned}$$

The collections of object methods and class methods are predefined for convenience:

$$\begin{aligned}
& \textbf{let}\ om = (\text{``}\textit{init}\text{''} \mapsto \textit{objinit}) \oplus (\text{``}\textit{dnu}\text{''} \mapsto \textit{objdnu}) \oplus (\text{``}\textit{gc}\text{''} \mapsto \textit{objgc}) \\
& \textbf{let}\ cm = (\text{``}\textit{init}\text{''} \mapsto \textit{classinit}) \oplus (\text{``}\textit{new}\text{''} \mapsto \textit{classnew})
\end{aligned}$$

The meta-circular definition of the ASMI classes is given below:

$$\begin{aligned}
\textbf{let}\ & \textit{object} = \textit{send}(\textit{class}, \text{``}\textit{new}\text{''}, [[], [\text{``}\textit{class}\text{''}], om]) \\
& \textit{class} = \textit{send}(\textit{class}, \text{``}\textit{new}\text{''}, [[\textit{object}], \textit{cdv}, \textit{cm}])
\end{aligned}$$

The classes are "bootstrapped" using the same reasoning as is discussed in §4.4, except ASMI is significantly simpler since it consists of two classes and not eight. Even though there are fewer classes, ASMI does not represent a reduction in expressive power, since AS can be implemented by defining a method and a pair of classes:

**let** $\textit{asnew}(e_1, l_1, l_2, e_2) = \textit{send}(\textit{send}(\textit{class}, \text{``}\textit{new}\text{''}, [\backslash_r(::)(f)([])(l_1), [], e_1]), \text{``}\textit{new}\text{''}, [l_1, l_2, e_2])$
  **where** $f(c) = \textit{send}(c, \text{``}\textit{gc}\text{''}, ())$

**let** $\textit{asm} = \textit{send}(\textit{class}, \text{``}\textit{new}\text{''}, [[\textit{class}], [], \text{``}\textit{new}\text{''} \mapsto \textit{asnew}])$

**let** $\textit{asc} = \textit{send}(\textit{asm}, \text{``}\textit{new}\text{''}, [[\textit{asm}], [], \{\}])$

# 6 ASMI with Reflective Send

§5 defines an OOPL called ASMI with multiple inheritance; this is a generalisation of AS which only supports single inheritance. It is not possible to define ASMI using the reflective facilities of AS since the inheritance strategy of AS depends upon *send* which is an intransigent language feature. This section defines the language Abstract Smalltalk with Multiple Inheritance and Reflective Send (ASMIRS) which extends the reflective power of ASMI by making *send* a method which is implemented at the meta-level. Since *send* is a method, it may be extended or redefined; this allows for programmer control of the inheritance strategy at the meta-level. Infinite regress is prevented because the language ASMIRS is meta-circular and the fixed point of $\mathcal{M}$ causes message passing to "bottom out".

§6.1 compares ASMIRS with ASMI and AS, §6.2 describes the ASMIRS basic classes, §6.3 extends the object representation from ASMI so that it becomes a reflective language feature, §6.4 describes message passing and §6.5 gives the meta-circular definition of ASMIRS.

## 6.1 Comparison with ASMI

ASMI represents objects as graphs where the nodes are instance variable and method environments and the multiple edges leading from a node arise due to multiple inheritance. The object creation and message delivery service require intimate knowledge of an object graph in order to construct them and extract methods with a given selector. There is no scope for representing objects in any other way or for making the message delivery service dependent upon the type of an object. Object representation and message delivery are intransigent language features in ASMI which means that there is only one possible object-graph traversal scheme when delivering a message.

ASMIRS allows the representation and message delivery service of an object to depend upon its type. Object creation is performed by sending a *new* message to a class. Message delivery is performed by sending a *send* message to the class of the receiver. Multiple object representations and message delivery services can co-exist within ASMIRS. ASMI is easily implemented in ASMIRS which will also allow different variations of the ASMI object-graph traversal scheme.

## 6.2 Basic Classes

ASMIRS is defined as a basic collection of classes whose initial configuration is shown in figure 2. §6.2.1 describes *object* and §6.2.2 describes *class*.

### 6.2.1 Object

$$\begin{aligned}
ienv = &(\text{``}supers\text{''} \mapsto [\,]) \oplus \\
&(\text{``}ivars\text{''} \mapsto [\text{``}class\text{''}]) \oplus \\
&(\text{``}menv\text{''} \mapsto (\text{``}init\text{''} \mapsto objinit) \oplus \\
&\qquad (\text{``}dnu\text{''} \mapsto objdnu)) \oplus \\
&(\text{``}class\text{''} \mapsto class)
\end{aligned}
\qquad
\begin{aligned}
menv = &(\text{``}new\text{''} \mapsto classnew) \oplus \\
&(\text{``}init\text{''} \mapsto classinit) \oplus \\
&(\text{``}on\text{''} \mapsto classon) \oplus \\
&(\text{``}send\text{''} \mapsto classsend) \oplus \\
&(\text{``}dnu\text{''} \mapsto objdnu)
\end{aligned}$$

*object* defines a single instance variable *class* whose value in any object is the class which was instantiated to produce the object and two methods for initialisation and handling unknown messages. *object* is an instance of *class*.

### 6.2.2 Class

$$ienv = (\text{``supers''} \mapsto [object]) \oplus$$
$$(\text{``ivars''} \mapsto [\text{``supers''}, \text{``ivars''}, \text{``menv''}]) \oplus$$
$$(\text{``menv''} \mapsto (\text{``init''} \mapsto classinit) \oplus$$
$$(\text{``new''} \mapsto classnew) \oplus$$
$$(\text{``on''} \mapsto classon) \oplus$$
$$(\text{``send''} \mapsto classsend)$$
$$(\text{``class''} \mapsto class)$$

$$menv = (\text{``new''} \mapsto classnew) \oplus$$
$$(\text{``init''} \mapsto classinit) \oplus$$
$$(\text{``on''} \mapsto classon) \oplus$$
$$(\text{``send''} \mapsto classsend) \oplus$$
$$(\text{``dnu''} \mapsto objdnu)$$

*class* defines three instance variables *supers*, *ivars* and *menv*. It defines four methods: *new* which is used to create objects and is implemented using *classnew* which represents objects as ASMI graphs, *init* which is used to initialise classes and is implemented by *classinit*, *on* which is used to order the nodes in an object-graph and is implemented by *classon*, and *send* which is the message delivery service for objects which are implemented as ASMI graphs and is implemented by *classsend*.

### 6.3 Representation

An ASMIRS object is represented as a value

$$obj(c, v)$$

where $c$ is the class which was instantiated to produce the object and $v$ is a value which is the objects internal representation. The functions *classof* and *repof* extract the class and representation from an object respectively.

An object is usually created by sending a class $c$ a *new* message. The class will respond to this message by constructing a representation for the object which is appropriate to the way that the class implements its message handling service, *i.e.* the method *send*. In this way, all objects have a uniform representation containing a class and a value but the value is manipulated at the meta-level by methods which are implemented by the class.

### 6.4 Message Passing

A message is sent to an object using the ASMIRS *send* primitive:

$$send(o, n, v)$$

The message is delivered to the object (and executed) by the delivery service which is implemented by the class of *o*. A delivery service is implemented by a method *send* and performed by sending the message:

$$send(classof(o), \text{``send''}, (o, n, v))$$

Since *classof(o)* is itself an object, the *send* message is performed by the delivery service which is implemented by its class:

$$send(classof(classof(o)), \text{``send''}, (classof(c), \text{``send''}, (o, n, v)))$$

When does this regression terminate? ASMIRS has an initial class configuration which is the same as that for ASMI shown in figure 2. Any (direct) instance of *class* will have a delivery service which is implemented by *classsend*. This knowledge is built into the basic message sending primitive. For any well-formed configuration of ASMIRS objects, a chain of *classof* calls will eventually lead to an instance of *class*, whose delivery service is known. The message delivery primitive is defined below:

$$\begin{aligned}
&\textbf{let }\ send(o,n,v) = \\
&\quad \textbf{if }\ classof(o) = class \\
&\quad \textbf{then case }\ n\ \textbf{of} \\
&\qquad\qquad \text{``}on\text{''} \Rightarrow classon(\epsilon,\epsilon,\epsilon,v) \\
&\qquad\qquad \text{``}send\text{''} \Rightarrow classsend(\epsilon,\epsilon,\epsilon,v) \\
&\qquad\qquad \_ \Rightarrow send(classof(o), \text{``}send\text{''}, (o,n,v)) \\
&\qquad \textbf{end} \\
&\quad \textbf{else }\ send(classof(o), \text{``}send\text{''}, (o,n,v))
\end{aligned}$$

It "knows" about very little of the ASMIRS system other than the minimal knowledge of how *class* is implemented. Other than this, all messages are delivered by the particular service depending upon the type of the receiver.

## 6.5 Definition

ASMIRS is defined as an initial collection of classes and a delivery service *send*. The classes define methods which are implemented using the functions given below:

$\textbf{let meth }\ classon(o) = onr(repof(o))$

$\textbf{let meth }\ classnew(c,v) = send(obj(c, instantiate(repof(c))), \text{``}init\text{''}, (c,v))$

$$\begin{aligned}
&\textbf{let meth }\ classsend(o,n,v) = \\
&\quad \textbf{let }\ x = find(p)(\epsilon)(send(classof(o), \text{``}on\text{''}, o))\ \textbf{ where }\ p(x) = n \in dom(getmeths(x)) \\
&\quad \textbf{in if }\ x = \epsilon \\
&\qquad \textbf{then }\ send(obj(classof(o), unmark(repof(o))), \text{``}dnu\text{''}, (n,v)) \\
&\qquad \textbf{else let }\ o_1 = \backslash_l(mark)(\mathbf{I})(1st(splitlistl(x, send(classof(o), \text{``}on\text{''}, o)))) \\
&\qquad\qquad\qquad o_2 = obj(classof(o), unmark(repof(o))) \\
&\qquad\qquad\qquad e = \backslash_r(\oplus)(getmeths)(\{\})(2nd(splitlistr(x, send(classof(o), \text{``}on\text{''}, o)))) \\
&\qquad\quad \textbf{in }\ ((getmeths(x)) \bullet n)(o_2, o_1, e)(v)
\end{aligned}$$

The methods defined by *class* are defined to be *cm*:

$$\begin{aligned}
\textbf{let }\ cm = &(\text{``}on\text{''} \mapsto classon) \oplus (\text{``}new\text{''} \mapsto classnew)\ \oplus \\
&(\text{``}send\text{''} \mapsto classsend) \oplus (\text{``}init\text{''} \mapsto classinit)
\end{aligned}$$

The meta-circular definition of ASMIRS classes is given below:

$$\begin{aligned}
\textbf{let }\ object =&\ send(class, \text{``}new\text{''}, [[],[\text{``}class\text{''}], om]) \\
class =&\ send(class, \text{``}new\text{''}, [[object], [\text{``}supers\text{''}, \text{``}ivars\text{''}, \text{``}menv\text{''}], cm])
\end{aligned}$$

The following class, *c* is an example of the reflective power of ASMIRS at work. *c* implements objects as ASMI graphs but uses *onl* to order the nodes rather than *onr*.

$\textbf{let meth }\ con(o) = onl(repof(o))$

$\textbf{let }\ c = send(class, \text{``}new\text{''}, [[class], [], \text{``}on\text{''} \mapsto con])$

# 7 Conclusion, Further and Related Work

This paper has described three object-oriented systems of increasing reflective power: AS, ASMI and ASMIRS. The initial system is based on the initial configuration of Smalltalk classes. AS implements single inheritance which is generalised to multiple inheritance in ASMI. There are a wide variety of multiple inheritance implementation strategies which cannot co-exist within ASMI since the object representation and message delivery service are intransigent language features. These restrictions are lifted in ASMIRS which allows multiple object implementations and message delivery services to co-exist within the same language by lifting them to the meta-level. The semantics of AS, ASMI amd ASMIRS is made precise by implementing them in a functional language which has been specifically designed with primitive features for object-oriented languages.

This work is closely related to the ideas of the ObjVLisp community [13] [5] [6] which greatly simplified the representation of reflective object-oriented programming features. See [20] [15] and [17] for more recent work on particlar aspects of metaclasses in object-oriented programming languages. The use of first class identifier binding environments as a basis for objects and related programming language features is described in [23] [1] [18] [11] [12] [22] and [29]. There is a collection of literature describing elegant record (environment) calculi used as the basis of object oriented programming language semantics including [8] [9] [10] and [21].

The issues of reflection in object-oriented programming languages has been developed from an abstract (although executable) point of view. ASMIRS is a simple but precise specification of an object-oriented programming language which has a very expressive reflective capability. ASMIRS can be used as the basis of experimentation for new reflective mechanisms. Further work is necessary to find efficient implementation techniques for the features which have been described in this paper. The reflective features which have been described are strongly related to a particular type of programming language, namely class based object-oriented languages. A functional programming language whose semantics is described in terms of a state transition system (such as the SECD machine) is particularly amenable to the *reification* of computational entities (such as environments) which are necessary to describe reflective features and allow a program to control its own execution. It would be interesting to see other reflective mechanisms could be designed using simple extensions to the underlying language.

# References

[1] Agha, G. *The Structure and Semantics of Actor Languages.* In Foundations of Object-Oriented Languages LNCS 489 (1990) 1 – 59.

[2] Bird, R. & Wadler P. *Introduction to Functional Programming.* Prentice Hall International Series in Computer Science. (1988)

[3] Bobrow D. *et al. Common Lisp Object System Specification.* Lisp and Symbolic Computation 1, 3/4. (Jan. 1989)

[4] Bobrow D. & Stefik M. *The Loops Manual.* Intelligent Systems Laboratory, Xerox PARC. (1983)

[5] Briot J-P. & Cointe P. *The ObjVLisp Model: Definition of a Uniform, Reflexive and Extensible object Oriented Language.* ECAI (1986)

[6] Briot J-P. & Cointe P. *A Uniform Model for object Oriented Languages Using the class Abstraction.* IJCAI (1987)

[7] Cardelli L. *The Semantics of Multiple Inheritance.* Proceedings of the Conference on the Semantics of Data Types. Springer-Verlag LNCS (June 1984) 51 – 66.

[8] Cardelli L. *A Semantics of Multiple Inheritance.* LNCS 173 The Semantics of Data Types. (1984)

[9] Cardelli L. & Mitchell J. *Operations in Records.* Math. Struct. In Comp. Science 1 (1991) 3 – 48

[10] Cardelli L. & Wegner P. *On Understanding Types, Data Abstraction and Polymorphism.* Computing Surveys 17, 4 (1985)

[11] Clark, A. N. *Semantic primitives for OOPLS.* Forthcoming PhD Thesis, QMW College, London University.

[12] Clark, A. N. *A Layered Object-Oriented Programming Language.* To appear early 1995 in The GEC Journal of Research. Also submitted (in a greatly extended form) to ACM TOPLAS, Oct. 1994.

[13] Cointe P. *Metaclasses are First Class: The ObjVLisp Model.* OOPSLA (1987)

[14] Ducournau R. et al. *Monotonic Conflict Resoloution Mechanisms for Inheritance.* OOPSLA (1992) 16 – 24

[15] Ferber J. *Computational Reflection in Class based Object-Oriented Languages.* OOPSLA (1989) 317 – 326

[16] Field, A. J. & Harrison P. G. *Functional Programming.* Addison Wesley International Computer Science Series. (1988)

[17] Foote B. & Johnson R. E. *Reflective Facilities in Smalltalk-80.* OOPSLA (1989) 327 – 335

[18] Gelernter D. *Environments as First Class Objects.* POPL 14.

[19] Goldberg A. & Robson D. *Smalltalk-80 The Language and its Implementation.* Addison-Wesley. (1983)

[20] Graube N. *Metaclass Compatibility.* OOPSLA (1989) 305 – 315

[21] Harper R. et al. *A Record Calculus Based on Symmetric Concatenation.* POPL 18.

[22] Jagannathan, S. Metalevel Building Blocks for Modular Systems. *ACM TOPLAS 16 3.* (1994)

[23] Jagannathan S. & Agha G. *A Reflective Model of Inheritance.* ECOOP (1992)

[24] Kiczales G. et al. *The Art of the Metaobject Protocol.* MIT Press (1991)

[25] Landin P. *The Mechanical Evaluation of Expressions.* Computer J. 6 (1964)

[26] Landin P. *The Next 700 Programming Languages.* Comm. ACM 9, 3 (1966)

[27] Maes P. *Reflection in an object Oriented Language.* Vrije Universiteit Brussel. AI Memo 86-8. (1986)

[28] Meyer B. *Eiffel The Language.* Prentice Hall object Oriented Series. (1992)

[29] Miller J. & Rozas G. *Free Variables and First-Class Environments.* Lisp and Symbolic Computation 4 (1991) 107 – 141

[30] Ossher H. & Harrison W. *Combination of Inheritance Hierarchies.* OOPSLA (1992)

[31] Peyton Jones S. *The Implementation of Functional Programming Languages.* Prentice Hall International (1987)

[32] Plotkin G. *Call-by-name, call-by-value and the $\lambda$-calculus.* Theoretical Computer Science 1. (1975)

[33] Snyder A. *Encapsulation and Inheritance in Object-Oriented Programming.* OOPSLA (1986)

[34] Steels L. *The KRS Concept System.* Vrije Universiteit Brussel. AI Lab. Technical Report 86-1 (1986)

[35] Wadler, P. *Comprehending Monads.* In Proc. $19^{th}$ Symp. on Lisp and Functional Programming. Nice. ACM. (1990)

## A  Functional Language Semantics

The kernel language syntax allows top level recursive definitions $T_0$ and expressions $E_0$:

$T_0 ::= \textbf{let } I = E_0$
$E_0 ::= I \mid N \mid S \mid \lambda I.E_0 \mid E_0 E_0 \mid \textbf{if } E_0 \textbf{ then } E_0 \textbf{ else } E_0 \mid (E_0, \ldots, e_0) \mid E_0; E_0$

The semantics of the kernel language is that of a simple call-by-value functional language. It is beyond the scope of this paper to fully describe the kenel semantics – see [11] for a complete description which is based upon Landin's SECD machine [25] extended with primitive features for object-oriented programming. The semantics of the sugared language is given as a translation to the kernel as follows: Let $\longrightarrow$ be a rewrite rule which removes the outer layer of sugar from an expression. $\longrightarrow$ is defined by case analysis below and, by repeated application to a sugared expression, will produce a kernel expression.

A curried function is translated by making all of the functions explicit

$$\lambda p_1 p_2 \ldots p_n.e \longrightarrow \lambda p_1.\lambda p_2.\ldots \lambda p_n.e$$

A function with a pattern parameter is translated to a function with a new parameter $i$ which returns a distinguished value $\epsilon$ if the supplied value does not match the pattern. A tuple pattern uses *isntuple* to test whether the supplied vaue is a tuple of the correct length and uses $\_\uparrow\_$ to extract the required components from the tuple

$$\lambda(p_1,\ldots,p_n).e \longrightarrow \lambda i.\ \textbf{if}\ isntuple(i)\ \textbf{then}\ \textbf{let}\ p_1 = i \uparrow 1 \ldots p_n = i \uparrow n\ \textbf{in}\ e\ \textbf{else}\ \epsilon$$

A constructor pattern uses *isk* and *stripk* to test whether the supplied value is constructed using the constructor $k$ and to strip the constructor to reveal the value underneath

$$\lambda kp.e \longrightarrow \lambda i.\ \textbf{if}\ isk(i)\ \textbf{then}\ \textbf{let}\ p = stripk(i)\ \textbf{in}\ e\ \textbf{else}\ \epsilon$$

A constant pattern tests whether or not the supplied value is the required constant

$$\lambda c.e \longrightarrow \lambda i.\ \textbf{if}\ i = c\ \textbf{then}\ e\ \textbf{else}\ \epsilon$$

An overloaded function definition is translated to a single function definition where the alternatives are composed using the infix operator ▌ which is defined as follows:

$$(f_1 ▌ f_2)(v) = \textbf{let}\ x = f_1(v)\ \textbf{in}\ \textbf{if}\ x = \epsilon\ \textbf{then}\ f_2(x)\ \textbf{else}\ x$$

Function definitions are translated using the operators *name* and *body*

$$f \to name(f) = body(f)$$

The name of a collection of function definitions is only defined when the names of the functions are all the same (note that $\tilde{p}$ denotes a sequence of patterns)

$$\begin{aligned}name(i\tilde{p} = e) &= i \\ name(\textbf{meth}\ i\tilde{p} = e) &= i \\ name(f_1 \mid f_2) &= name(f_1)?name(f_2)\end{aligned}$$

where ? combines identifiers, returning one of the identifiers if they are both the same and is undefined otherwise. The operator *body* translates a named function or method to an anonymous function or method

$$\begin{aligned}body(i\tilde{p} = e) &= \lambda \tilde{p}.e \\ body(\textbf{meth}\ i\tilde{p} = e) &= \textbf{meth}\ \tilde{p}.e \\ body(f_1 \mid f_2) &= body(f_1) ▌ body(f_2)\end{aligned}$$

A method is a function which has some hidden parameters

$$\textbf{meth}\ \tilde{p}.e \longrightarrow \lambda(self, next, i)\ \tilde{p}.\ \textbf{open}\ i\ \textbf{in}\ e$$

The identifiers named *self* and *next* are scoped over the body of the method $e$ and are named *self* and *super* in Smalltalk. The identifier $i$ is bound to an instance variable environment which is opened for the scope of the method body. It is important that this identifier does not capture any free references to identifiers in $e$. In the presence of environment reification and installation this is not achieved simply through a static analysis of the code.
All infix operators are curried

$$e_1 o e_2 \to (o(e_1))(e_2)$$

Sequences of declarations are translated to a single declaration

$$i_1 = e_1\ i_2 = e_2\ \ldots\ i_n = e_n \longrightarrow (i_1, i_2, \ldots, i_n) = (e_1, e_2, \ldots, e_n)$$

Let and where expressions are translated to function applications

$$e_1 \text{ where } p = e_2 \longrightarrow (\lambda p.e_1)(e_2)$$

$$\text{let } p = e_1 \text{ in } e_2 \longrightarrow (\lambda p.e_2)(e_1)$$

A case expression is translated to a function application

$$\text{case } e \text{ of } a \text{ end} \longrightarrow a(e)$$

A case arm is translated to a partial function

$$p \to e \longrightarrow \lambda p.e$$

$$a_1 a_2 \longrightarrow a_1 \,|\, a_2$$

An open expression is translated to use the builtin primitives for environment manipulation

$$\text{open } e_1 \text{ in } e_2 \longrightarrow (e_1 \hookrightarrow \lambda().e_2)()$$